\documentclass[a4paper,11pt]{article}
\usepackage{jinstpub}

\proceeding{28$^{\text{th}}$ International Workshop on Radiation Imaging Detectors\\
28 June -- 2 July 2026\\
Ghent, Belgium}

\title{\boldmath Irradiation Studies and Design Optimization of the ATLAS Tile Calorimeter for the High-Luminosity LHC}

\author{R\'obert Astalo\v{s}}
\affiliation{Comenius University Bratislava,\\ Bratislava, Slovakia}
\collaboration[c]{on behalf of the ATLAS Tile Calorimeter System}

\emailAdd{robert.astalos@cern.ch} 

\abstract{The Tile Calorimeter (TileCal) is a sampling hadronic calorimeter covering the central region of the ATLAS experiment at the CERN Large Hadron Collider (LHC). It employs steel as the absorber material and plastic scintillators as the active medium. The High-Luminosity LHC (HL-LHC), scheduled to start operation in 2030, will deliver instantaneous luminosities significantly exceeding the baseline LHC design, imposing more stringent requirements on detector readout and trigger systems. Consequently, TileCal has to be capable of reliable operation under increased radiation levels and very high particle flux, while maintaining full compatibility with the upgraded ATLAS trigger architecture.
During the Long Shutdown period (2026–2030), the TileCal readout electronics will be entirely replaced with radiation-tolerant systems designed to handle data rates approximately an order of magnitude higher than those of the baseline LHC configuration. The photomultiplier tubes (PMTs) in the most highly irradiated regions will also be replaced with improved devices exhibiting enhanced stability and radiation tolerance.
To meet these challenges, the system design has been correspondingly optimized to ensure improved performance, efficiency and robustness in high-radiation environments, and an extensive irradiation testing program has been carried out. This contribution presents the resulting design developments of the TileCal system, together with the results obtained from the irradiation tests.}


\begin{document}
\begingroup
\renewcommand\thefootnote{}
\footnotetext{Copyright 2026 CERN for the benefit of the ATLAS Collaboration. CC-BY-4.0 license.}
\endgroup
\maketitle
\flushbottom

\section{Introduction}
\label{sec:intro}
\begin{figure}[htbp]
\centering
\includegraphics[width=.52\textwidth]{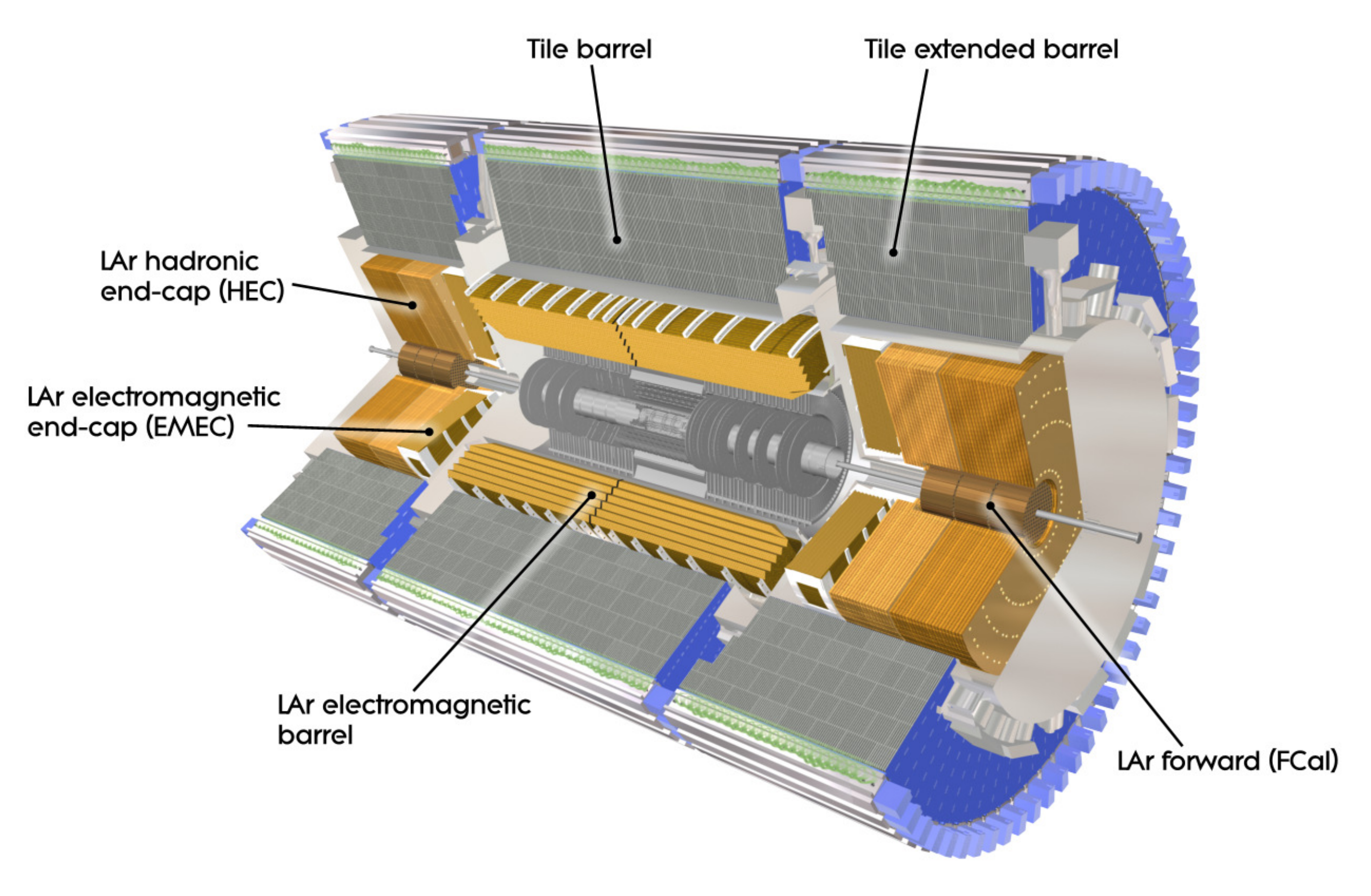}
\qquad
\includegraphics[width=.26\textwidth]{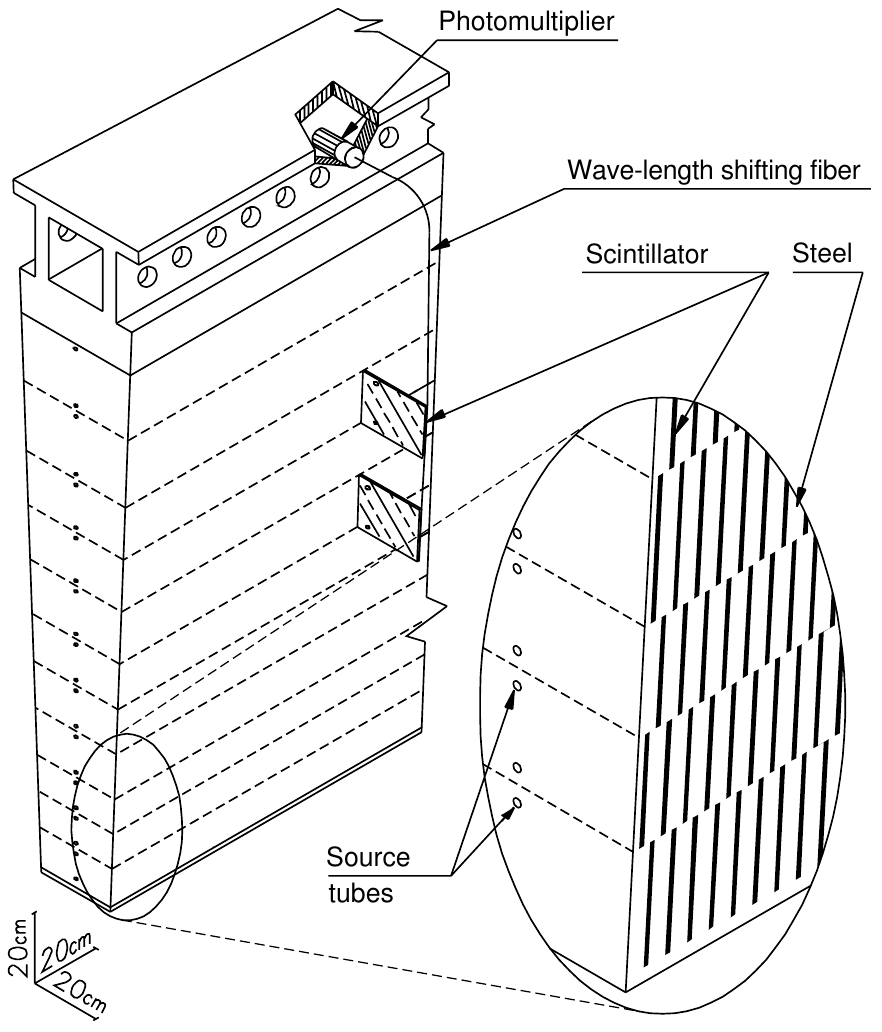}
\caption{The ATLAS Tile Calorimeter (left) and the internal structure of a wedge-shaped module slice (right)~\cite{atlas_tdr}.\label{fig:tilecal_diagram}}
\end{figure}

\vspace{-0.4em}
ATLAS is a multi-purpose particle physics experiment located at the CERN Large Hadron Collider (LHC)~\cite{atlas_experiment}. Over the next decade, the LHC will undergo a major upgrade to become the High-Luminosity LHC (HL-LHC). This upgrade is designed to significantly increase the collision rate, aiming to deliver an ultimate integrated luminosity of up to 4000\,fb$^{-1}$. While this vast amount of data will open new frontiers in high-energy physics, it also introduces a significantly more demanding radiation environment.
The ATLAS Hadronic Tile Calorimeter (TileCal) is a critical component for measuring hadron and jet energies, hadronic $\tau$-decays, and missing transverse energy. TileCal is a sampling hadronic calorimeter consisting of wedge-shaped modules. These modules are arranged into a central Long Barrel (LB) and two Extended Barrels (EB). In total, the calorimeter contains 256 modules and is segmented into 5182 calorimetric cells. The active medium of the detector consists of scintillating plastic tiles, while steel is used as the absorber material. Ionizing particles traversing the plastic tiles produce scintillation light, which is collected and shifted to longer wavelengths by wavelength-shifting fibers. This light is then channeled to photomultiplier tubes (PMTs) located at the outer radius of the modules. To ensure redundancy and uniform response, most cells are read out by two PMTs, with the exception of the special E1--E4 gap/crack cells, which are read out by a single PMT. A macroscopic view of the ATLAS Tile Calorimeter and a wedge-shaped module slice are presented in figure~\ref{fig:tilecal_diagram}.

Due to the increased instantaneous and integrated luminosity of the HL-LHC, components exposed to the highest radiation levels in TileCal must tolerate significantly elevated levels of ionizing and non-ionizing radiation. This is particularly crucial for the Long Barrel Low Voltage Power Supplies (LVPS) and the readout electronics situated close to the detector gap regions. Consequently, an entirely new on-detector architecture has been designed.

\section{TileCal HL-LHC On-Detector Architecture}
\label{sec:architecture}
\vspace{-0.4em}
The upcoming Phase-II upgrade entails a complete replacement of the on-detector readout electronics to cope with the increased data rates and radiation levels. The newly designed architecture is highly modular and incorporates modern radiation-tolerant commercial off-the-shelf (COTS) components alongside application-specific integrated circuits (ASICs).

\begin{figure}[htbp]
\centering
\includegraphics[width=0.65\textwidth]{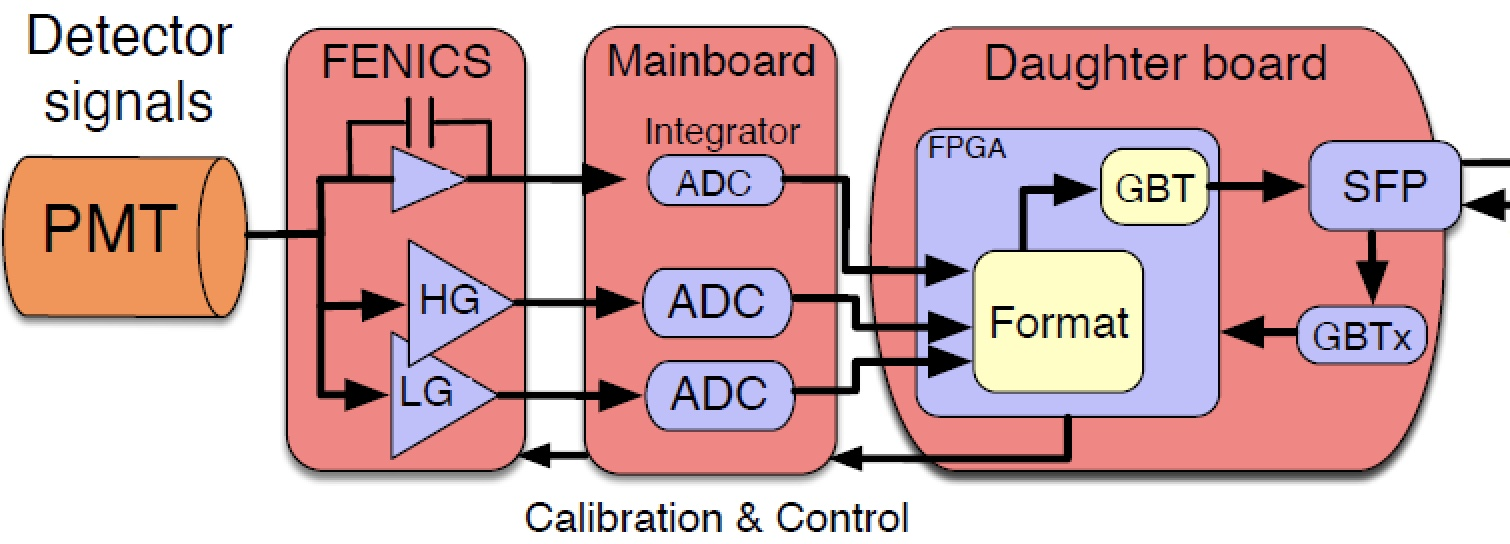}
\caption{The Tile HL-LHC On-Detector architecture block diagram, detailing the connectivity between power distribution, front-end signal processing, and data transmission~\cite{atlas_tdr}.\label{fig:architecture}}
\end{figure}

The block diagram of the Tile HL-LHC on-detector architecture is shown in figure~\ref{fig:architecture}.

The architecture is composed of several critical sub-systems:
\vspace{-0.8em}
\begin{itemize}
    \item \textbf{LVPS bricks:} A total of 1792 LVPS bricks utilize switching DC-DC converters operating at 300\,kHz. They step down a bulk 200\,V input to distribute stable, low-voltage power to the sensitive on-detector readout electronics. 
    \vspace{-0.8em}
    \item \textbf{PMTs:} Approximately 1000 new PMTs (Hamamatsu R11187) will be installed in the most exposed regions, with a quantum efficiency exceeding 15\% and enhanced radiation tolerance.
    \item \textbf{Active Dividers:} The system requires 9852 Active Dividers to provide stable high voltage division for the PMTs. Their active regulation minimizes gain variations even at the extremely high event rates expected at the HL-LHC. 
    \vspace{-0.8em}
    \item \textbf{FENICS Cards:} A matching set of 9852 Front End board for the New Infrastructure with Calibration and signal Shaping (FENICS) cards are responsible for PMT pulse shaping with bi-gain amplification (1:40). They also feature a current integrator for continuous luminosity measurements and $^{137}$Cs calibration.
    \vspace{-0.8em}
    \item \textbf{MainBoards (MBs):} There will be 896 MainBoards serving as the central hubs for module control and configuration. They perform the digitization of the FENICS signals utilizing two 12-bit Analog-to-Digital Converters (ADCs) running at 40\,Msps, and one 16-bit ADC for the integrator readout.
    \vspace{-0.8em}
    \item \textbf{DaughterBoards (DBs):} Also totaling 896 units, the DaughterBoards function as the central communication hub of each module. They are responsible for transmitting approximately 35\,Tbps of digitized physics data to off-detector systems via 3584 optical uplinks operating at 9.6\,Gbps. Simultaneously, they receive critical configuration and timing commands via 1792 downlinks at 4.8\,Gbps.
\end{itemize}

\section{Radiation Criteria and Expected Doses}
\label{sec:radiation_criteria}
\vspace{-0.4em}
To ensure survivability and reliable operation throughout the HL-LHC lifespan, all on-detector electronics must be evaluated against three primary radiation damage mechanisms: Total Ionizing Dose (TID), which causes cumulative parametric shifts in semiconductor devices; Non-Ionizing Energy Loss (NIEL), which produces displacement damage in silicon lattices; and Single Event Effects (SEE), of which Single Event Upsets (SEU, non-destructive bit-flips) and Single Event Latch-ups (SEL, potentially destructive high-current shorts) are of primary concern.

Detailed simulations of the ATLAS radiation environment were performed to determine the expected doses for an integrated luminosity of 4000\,fb$^{-1}$. The simulated worst-case doses (without safety factors) for the highest irradiated regions are summarized in table~\ref{tab:doses} (fluences are denoted in 1\,MeV neutron equivalent per square centimeter, n/cm$^2$, and protons per square centimeter, p/cm$^2$). 

\begin{table}[htbp]
\centering
\caption{Simulated worst-case radiation doses for 4000\,fb$^{-1}$ (no safety factors included). Only the highest doses from the Barrel and Endcap regions are shown~\cite{atlas_tdr}.\label{tab:doses}}
\smallskip
\begin{tabular}{lccc}
\hline
\multicolumn{1}{c}{Component} & \multicolumn{3}{c}{Dose, fitted in region of max} \\
\cline{2-4}
 & TID [Gy] & NIEL [n/cm$^2$] & SEE [p/cm$^2$] \\
\hline
PMT Divider - Barrel  & 17.6 & $2.0\times 10^{12}$ & $1.9\times 10^{11}$ \\
FENICS - Barrel       & 10.6 & $1.6\times 10^{12}$ & $1.8\times 10^{11}$ \\
MB - Barrel           & 10.0 & $1.4\times 10^{12}$ & $9.2\times 10^{10}$ \\
LVPS - Barrel         & 53.6 & $3.5\times 10^{12}$ & $5.3\times 10^{11}$ \\
DB - Endcap           & 6.8  & $9.7\times 10^{11}$ & $4.9\times 10^{10}$ \\
\hline
\end{tabular}
\end{table}

During the qualification process, rigorous safety factors were applied to these simulated base values in accordance with internal ATLAS guidelines, establishing the final qualification thresholds for all approved components (e.g., 108\,Gy TID and $13.16\times 10^{12}$\,n/cm$^2$ NIEL for the DaughterBoards).

\section{Front-End Electronics Qualification}
\label{sec:frontend}
\vspace{-0.4em}
The front-end electronics directly connected to the PMTs—specifically the FENICS boards and Active Dividers—were subjected to extensive irradiation campaigns to validate their performance.

\subsection{FENICS Boards}
\begin{figure}[htbp]
\centering
\includegraphics[width=0.57\textwidth]{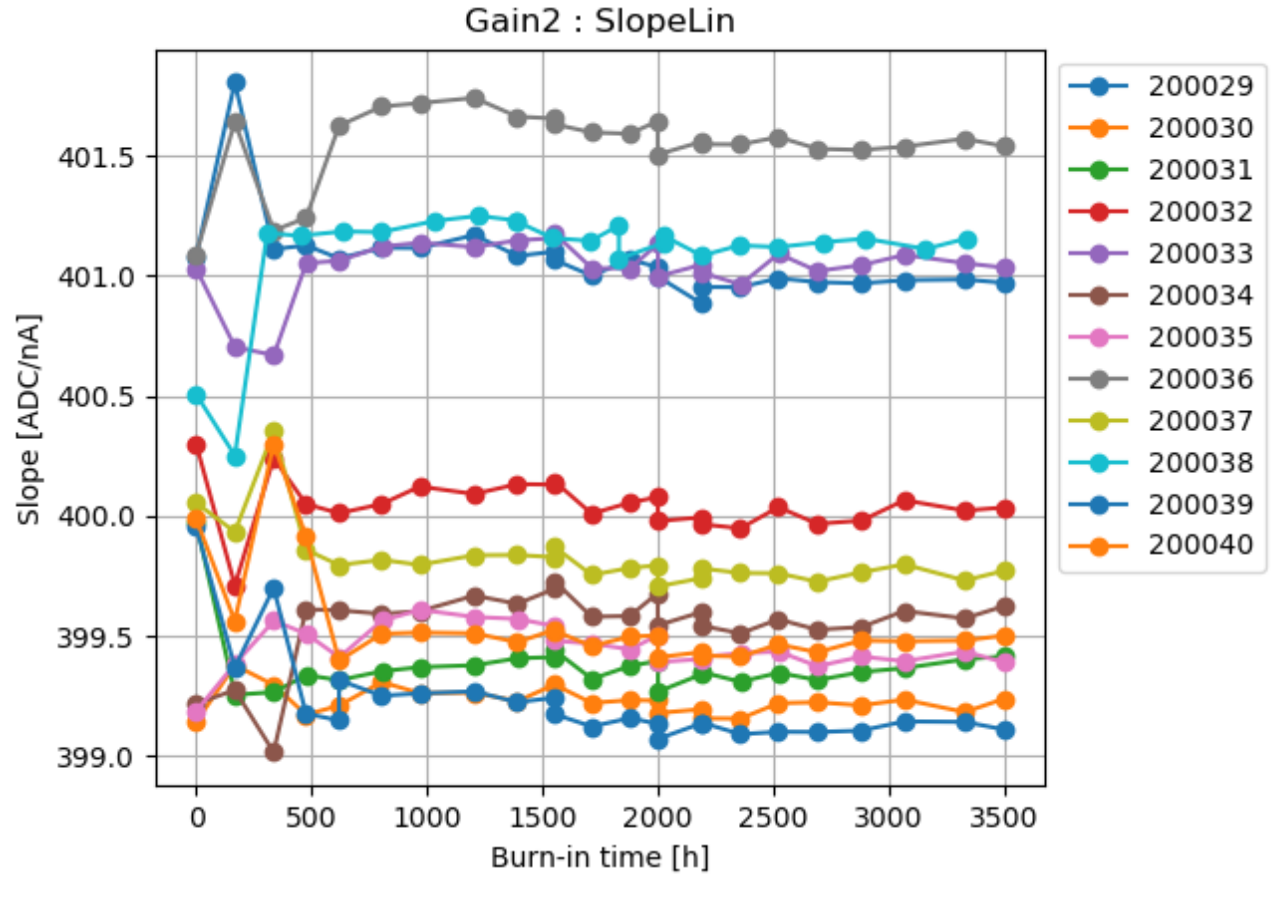}
\caption{Evolution of the FENICS integrator Gain2 as a function of burn-in time during radiation testing. Each line represents an individual readout channel, demonstrating the uniformity of the response across the board.\label{fig:fenics_noise}}
\end{figure}

\vspace{-0.4em}
Over 40 FENICS boards, in addition to pre-production tracking models, were irradiated with gamma rays and neutrons. The test campaigns evaluated TID limits up to 51\,Gy and NIEL limits up to $4.7 \times 10^{12}$\,n/cm$^2$. During these tests, the fast signal channels demonstrated exceptional resilience, showing no observable performance degradation. The integrator channels exhibited a slight, yet fully acceptable, variation in the Gain2 parameter over the burn-in time, as illustrated in figure~\ref{fig:fenics_noise}. This confirmed that all active COTS components selected for the FENICS boards are officially certified for production.

\subsection{Active Dividers}
\begin{figure}[htbp]
\centering
\includegraphics[width=0.85\textwidth]{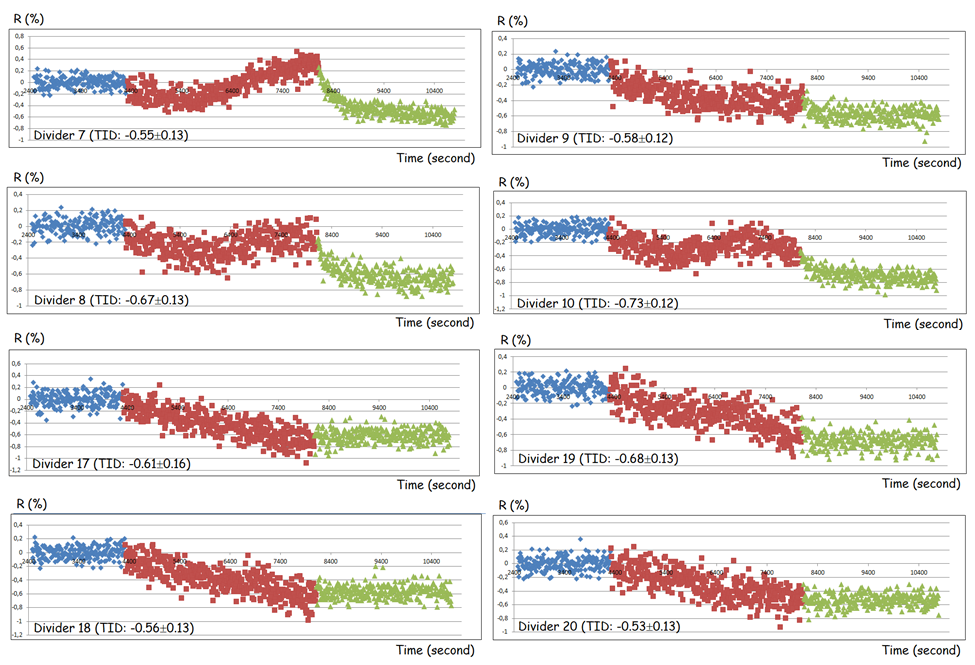}
\caption{Relative TID effects on the Photocathode-first dynode (Ph-D1) voltage of the Active Divider channels up to 480\,Gy. The blue, red, and green colors indicate measurements taken before, during, and after radiation exposure, respectively.\label{fig:divider_gain}}
\end{figure}

\vspace{-0.4em}
The Active Dividers underwent even more severe qualification tests, targeting 424\,Gy of TID, $1.37 \times 10^{13}$\,n/cm$^2$ of NIEL, and $2.97 \times 10^{12}$\,p/cm$^2$ for SEE evaluation. Through extensive proton, gamma, and neutron beam tests, it was observed that no Single Event Effects occurred during irradiation. Furthermore, the relative PMT gain decrease attributable to TID effects remained strictly within operational tolerances. At an absorbed dose of 480\,Gy, the average PMT gain deviation was measured at just $-1.92 \pm 0.20\%$. This robust performance is driven by the stability of the Photocathode-first dynode (Ph-D1) voltage; as shown in figure~\ref{fig:divider_gain}, the relative Ph-D1 voltage decrease across the tested divider channels was minimal, with an average deviation of just $-0.59 \pm 0.04\%$.

\section{On-Detector Readout Electronics Qualification}
\label{sec:readout}
\vspace{-0.4em}
The centralized digital processing and communication hubs of the module—the DaughterBoards and MainBoards—rely heavily on FPGAs and ADCs, which are inherently susceptible to radiation damage. 

\subsection{DaughterBoards}
\begin{figure}[htbp]
\centering
\includegraphics[width=0.48\textwidth]{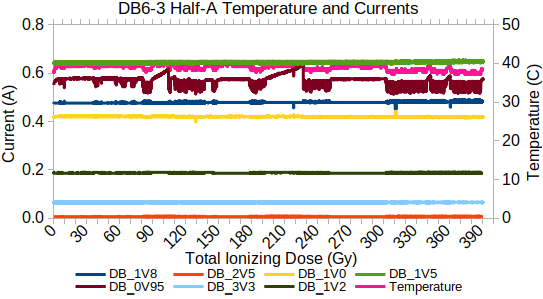}
\hfill
\includegraphics[width=0.48\textwidth]{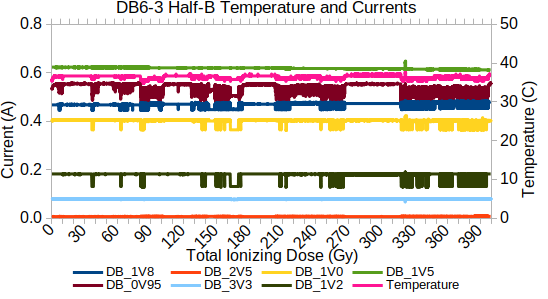}
\caption{Monitored KU FPGA temperatures and currents during 54\,MeV proton beam TID testing. The left and right plots correspond to the two redundant, symmetrical halves of the DaughterBoard (Half-A and Half-B, respectively). Sudden drops correspond to active SEU corrections by the Xilinx SEM and resets applied during uncorrectable SEUs~\cite{valdes_jinst}.\label{fig:db_tid}}
\end{figure}

\vspace{-0.4em}
The qualification criteria for the DaughterBoards required survival up to 108\,Gy TID and $13.16 \times 10^{12}$\,n/cm$^2$ NIEL. The DBs utilize Kintex Ultrascale (KU) and ProASIC FPGAs, which successfully withstood doses exceeding 108\,Gy and fluences of $14 \times 10^{12}$\,n/cm$^2$. Crucially, no destructive Single Event Latch-ups (SEL) were observed in the selected KU FPGAs. While Single Event Upsets (SEU) did occur in the logic fabric, these non-destructive soft errors were effectively mitigated on-the-fly without data loss by the Xilinx Soft Error Mitigation (SEM) IP core and Triple Mode Redundancy (TMR). Figure~\ref{fig:db_tid} displays the monitored temperature and current of the KU FPGA during a 54\,MeV proton beam TID test. The sudden drops in current prominently illustrate the active SEU corrections performed by the SEM and the necessary resets applied when uncorrectable multi-bit SEUs were encountered.

\subsection{MainBoards}
\vspace{-0.4em}
Similar rigorous testing was applied to the MainBoards. All components passed TID testing up to 640\,Gy and NIEL testing up to $1.4 \times 10^{13}$\,n/cm$^2$ without exhibiting any performance degradation. Critical components, including fast ADCs and FPGAs, showed zero SEU effects up to a fluence of $10^{13}$\,p/cm$^2$ during 800\,MeV proton testing, meaning they maintained full, uninterrupted functionality. Additionally, the Point-of-Load regulators passed stringent testing at 570\,Gy TID and $1.2 \times 10^{13}$\,n/cm$^2$ NIEL, securing the robustness of the local power delivery network.

\section{Low Voltage Power Supply Optimization and Testing}
\label{sec:lvps}
\vspace{-0.4em}
The Low Voltage power supply system was subjected to both severe radiation testing and a significant design optimization process to reduce its heat dissipation.

The LVPS architecture features 8 identical bricks for LB modules and 6 bricks for EB modules inside a shielded box. These bricks convert a bulk 200\,V input to a 10\,V output, operating at a nominal power of 23\,W. To preserve the TileCal dual-PMT cell redundancy, each brick powers an independent side of the readout electronics. 

A major design upgrade involved replacing the original STB57N65M5 power MOSFETs with SIHFS9N60A models. This change substantially reduced both the gate-source capacitance and the switching losses in the DC-DC converters. As a result, the power efficiency of the bricks was improved from 58\% to 72\%. This significant reduction in dissipated heat is essential for meeting the strict 77\,kW cooling system capacity limit of the TileCal detector.

The individual components of the optimized LVPS bricks were methodically irradiated. A summary of the SEE test conditions and the measured cross-sections is provided in table~\ref{tab:see_lvps}. The SEE cross section represents the probability of an event occurring per incident particle. 

\begin{table}[htbp]
\centering
\caption{Summary of Single Event Effect (SEE) test conditions and results for different LVPS components~\cite{moayedi_paper}.\label{tab:see_lvps}}
\smallskip
\begin{tabular}{lcccc}
\hline
\textbf{Component} & \textbf{Flux [p/cm$^2$/s]} & \textbf{Fluence [p/cm$^2$]} & \textbf{SEE Count} & \textbf{Cross Section [cm$^2$]} \\
\hline
LT1681 & $2.4 \times 10^8$ & $7.05 \times 10^{11}$ & 64 (2 chips) & $4.54 \times 10^{-11}$ \\
LTC6241 & $2.4 \times 10^8$ & $7.05 \times 10^{11}$ & 133 (4 op-amps) & $4.72 \times 10^{-11}$ \\
LT3080 & $2.4 \times 10^8$ & $7.05 \times 10^{11}$ & 198 (4 chips) & $7.02 \times 10^{-11}$ \\
IR2110 & $3.5 \times 10^8$ & $3.4 \times 10^{11}$ & None (2 chips) & $<1.47 \times 10^{-12}$ \\
SIHFS9N60A & $3.5 \times 10^8$ & $3.4 \times 10^{11}$ & None (8 chips) & $<3.67 \times 10^{-13}$ \\
SI8920 & $1 \times 10^7$ & $1.96 \times 10^{11}$ & 1 (20 chips) & $2.55 \times 10^{-13}$ \\
\hline
\end{tabular}
\end{table}

\begin{figure}[htbp]
\centering
\includegraphics[width=0.46\textwidth]{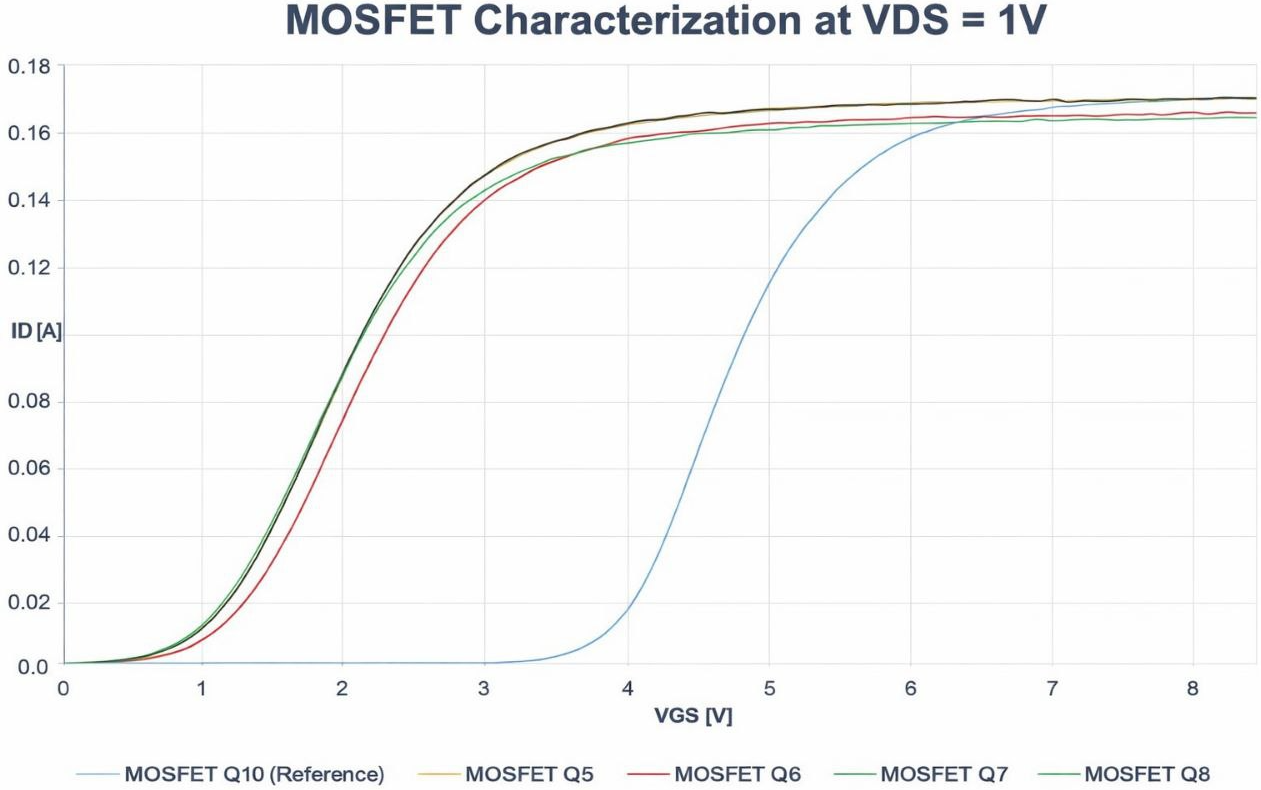}
\qquad
\includegraphics[width=0.46\textwidth]{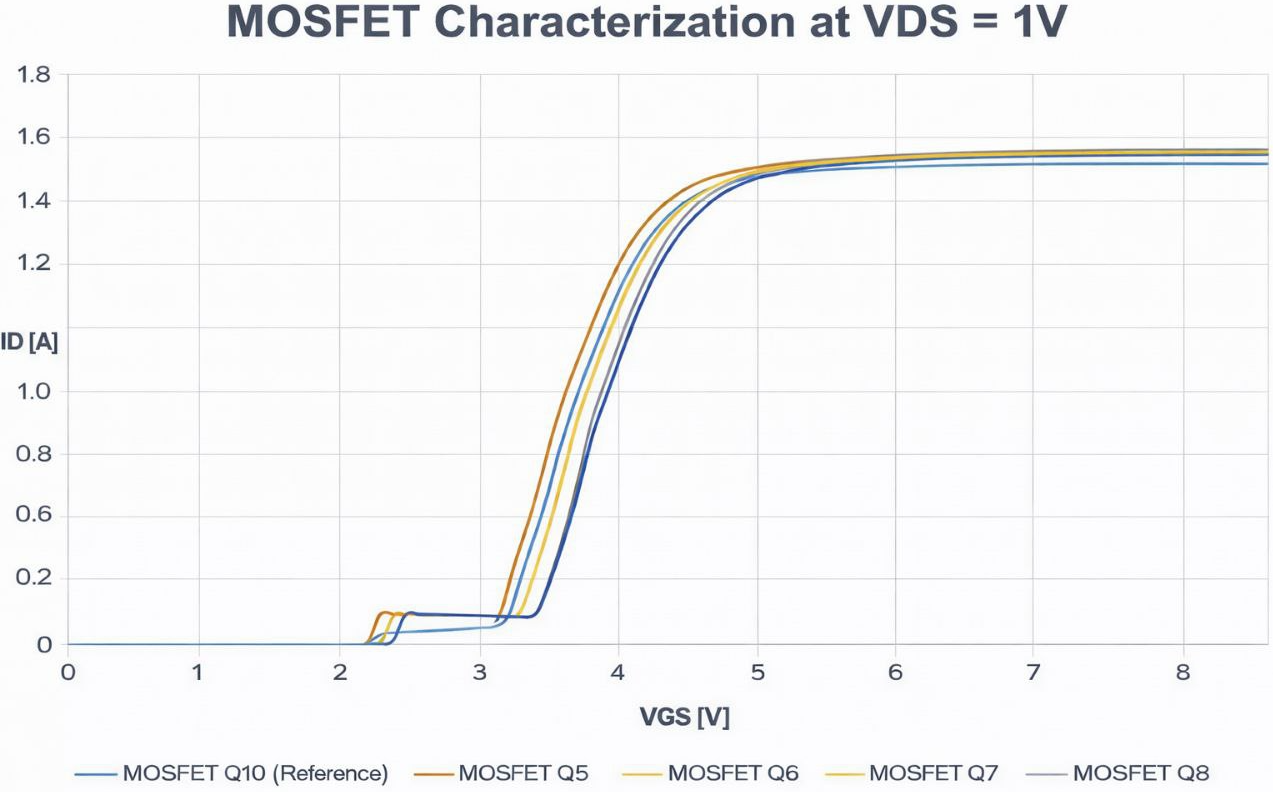}
\caption{SIHFS9N60A MOSFET characterization curves at $V_\mathrm{DS} = 1\,$V after absorbing 200\,Gy of TID (left) and after NIEL testing (right)~\cite{moayedi_paper}.\label{fig:mosfet}}
\end{figure}

The main high-power switching transistor, the SIHFS9N60A MOSFET, demonstrated excellent tolerance. A minor 2.5\,V shift in the gate-source threshold was observed after 200\,Gy of TID, and minimal deviation occurred following NIEL tests, as visualized in figure~\ref{fig:mosfet}.

The SI8920 Isolation Amplifier, which provides safe analog signal isolation for over-voltage protection, showed initial voltage fluctuations when driven at 150\,mV during tests. However, its performance proved to be completely stable below 100\,mV. This threshold was chosen to prevent false triggers in the over-voltage protection circuitry. Consequently, it has been selected for final production alongside specific circuitry designed to strictly limit its input voltage~\cite{moayedi_paper,moayedi_slide}. Furthermore, the LT1681 Controller and LT3080 Regulator underwent TID tests at a dose rate of 502\,Gy/h. They exhibited only minor output voltage drifts ($\leq 5\%$), which remain well within the acceptable operational limits of the power distribution network.

\section{Conclusions}
\label{sec:conclusion}
\vspace{-0.4em}
The ATLAS Tile Calorimeter readout electronics are undergoing a complete replacement to meet the unprecedented demands of the High-Luminosity LHC. The design optimization of the power distribution scheme has successfully enhanced readout reliability and significantly improved power efficiency from 58\% to 72\%, while effectively mitigating the impact of localized failures by utilizing identical, independent LVPS bricks.

Extensive irradiation campaigns have rigorously validated all design choices. All critical components for the upgrade of the TileCal on-detector electronics have successfully passed their target TID, NIEL, and SEE qualification thresholds. Having met or exceeded the stringent ATLAS radiation requirements, these components are fully certified and ready for mass production, ensuring the continued high-performance operation of the Tile Calorimeter throughout the HL-LHC era.


\begin{thebibliography}{99}
\vspace{-0.4em}
\bibitem{atlas_experiment}
ATLAS Collaboration,
\emph{The ATLAS Experiment at the CERN Large Hadron Collider},
\emph{JINST} {\bf 3} (2008) S08003.
\bibitem{atlas_tdr}
ATLAS collaboration,
\emph{Technical Design Report for the Phase-II Upgrade of the ATLAS Tile Calorimeter},
CERN-LHCC-2017-019 (2017).
\bibitem{moayedi_paper}
S.\ Moayedi,
\emph{Upgrade of the ATLAS Tile Calorimeter front-end power supply for the HL-LHC},
\emph{JINST} {\bf 21} (2026) C05002.
\bibitem{moayedi_slide}
S.\ Moayedi,
\emph{Upgrade of the ATLAS Tile Calorimeter Front-End Power Supply for the HL-LHC},
ATL-TILECAL-SLIDE-2025-552.
\bibitem{valdes_jinst}
E.\ Valdes Santurio et al.,
\emph{Radiation studies performed on the High Luminosity ATLAS TileCal link Daughterboard},
\emph{JINST} {\bf 18} (2023) C04011.
\end{thebibliography}
\end{document}